\documentclass[pre]{revtex4}

\usepackage{epsfig,amsmath,amssymb,graphics,color,calc}

\usepackage{graphicx}

\usepackage{amsmath}
\usepackage{amsthm}
\usepackage{amssymb}
\usepackage{amsopn}
\begin{document}
\title{The Student ensemble of correlation matrices: eigenvalue spectrum
and Kullback-Leibler entropy}

\author{Giulio Biroli$^{\dagger,*}$, Jean-Philippe Bouchaud$^{*}$, Marc
Potters$^*$}
\affiliation{
{$^\dagger$ Service de Physique Th{\'e}orique,
Orme des Merisiers -- CEA Saclay, 91191 Gif sur Yvette Cedex, France.}\\
{$^*$ Science \& Finance, Capital Fund Management, 6 Bd Haussmann,}
{75009 Paris France}
}

\newcommand \be  {\begin{equation}}
\newcommand \bea {\begin{eqnarray} \nonumber }
\newcommand \ee  {\end{equation}}
\newcommand \eea {\end{eqnarray}}
\newcommand{\subs}[1]{{\mbox{\scriptsize #1}}}
\newcommand \Tr {\mbox{Tr}}
\newcommand \rmi {\mbox{i}}
\newcommand \rmdd {\mbox{d}}

\begin{abstract}
We study a new ensemble of random correlation matrices related to multivariate Student (or more generally elliptic) random variables. 
We establish the exact density of states of empirical correlation matrices that generalizes
the Mar\v{c}enko-Pastur result. The comparison between the theoretical density of states in the Student case and empirical financial data is 
surprisingly good, even if we are still able to detect systematic deviations. Finally, we compute explicitely the Kullback-Leibler entropies 
of empirical Student matrices, which are found to be independent of the true correlation matrix, as in the Gaussian case. We provide numerically
exact values for these Kullback-Leibler entropies.
\end{abstract}

\maketitle

\section{Introduction}
Estimating and analyzing the correlation matrix of $N$ different 
variables from a data set is very recurrent 
problem in statistical analysis.  
Typically, one observes the value of the $N$ different variables 
over a time period of size $T$.  The total number of data
point is $NT$ whereas the number of elements of the correlation matrix is
$N(N-1)/2$. In many applications, both $T$ and $N$ are
large but $Q=T/N$ is of order unity. For example, in financial applications 
the values of $T$ and $N$ go from few hundreds to a few thousands.
In these cases, the correlation (or covariance) matrix is in fact 
rather poorly determined. 
This means, in particular, substantial noise in the determination of the
eigenvalues and eigenvectors.
For instance, focus on the case where all random variables  
are in fact independent, such that the true correlation matrix
$C_{ij}$ is the identity matrix. The {\it empirical} determination
of ${\bf C}$, which we call ${\bf E}$, is obtained from the zero mean, unit
variance Gaussian variables $r_i^t$ using the Pearson estimator:
\be\label{defE}
E_{ij}=\frac{1}{T}\sum_{t}^T r^t_i r^t_j
\ee
The eigenvalue spectrum $\rho(\lambda)$ of ${\bf E}$ only approaches a
$\delta$-function at $\lambda=1$ when $Q \to \infty$. When $Q$ is finite,
however,
$\rho(\lambda)$ is non-trivial. Under mild hypothesis about the
distribution of the $r_t^i$s (essentially that the variance is finite), the
spectrum
approaches, in the large $N,T$ limit, the Mar\v{c}enko-Pastur distribution
\cite{marcenkopastur}:
\be
\rho(\lambda) = \frac{\sqrt{4\lambda q - (\lambda-1+q)^2}}{2\pi \lambda q}
\qquad \lambda \in [(1-\sqrt{q})^2,(1+\sqrt{q})^2] \quad (q = 1/Q < 1).
\ee
For finite $N$, the edges of the spectrum are smoothed. The statistics of
exceptionally large eigenvalues is described by the Tracy-Widom
distribution \cite{GBA}, or by a Fr\'echet distribution when the $r_t$ have power-law
tails decaying sufficiently slowly \cite{BBP}. The probability of large fluctuations 
of the maximum eigenvalue (for Gaussian $r_t$) has been derived in \cite{VMB}.

In this work we focus on a more general ensemble that often provides a 
faithful representation of empirical data. We will consider that the random variables 
$r_i^t$ can be written as a product of two terms: $r_i^t=\sigma^t \eta_i^t$
where $\eta_i^t$ and $\sigma^t$ are independent random variables. 
The $\eta_i^t$ are characterized by a true correlation matrix 
$C_{ij}=\langle \eta_i \eta_j\rangle$ and 
$\sigma_i^t$ are such that $\langle \sigma^2 \rangle=1$. Furthermore, 
$\eta_i^t$ are drawn independently from the same distribution at each time $t$. 
A particular case on which we shall focus extensively corresponds to 
Gaussian distributed $\eta_i$ and
\be \label{defstudent}
P(\sigma)= \frac{2}{\Gamma(\frac{\mu}{2})}
\exp\left[-\frac{\sigma_0^2}{\sigma^2}\right] \frac{\sigma_0^\mu}{\sigma^{1+\mu}},
\ee
where $\sigma_0^2=2\mu/(\mu-2)$ in such a way that $\langle \sigma^2 \rangle=1$ ($\Gamma(x)$ is the Gamma function).
This case corresponds to a {\it multivariate Student distribution} for the original 
variables $r_i^t$. This defines the {\it Student Ensemble} 
we will consider more specifically below, although many of our results extend to 
arbitrary choice of $P(\sigma)$ and $P(\{\eta_i^t \})$, 
defining a large class of {\it elliptic} distributions. We will also focus on 
applications of our results to finance where the correct determination of the 
true correlation matrix between stocks play a very important role.

The application of Random Matrix Theory to financial correlation matrices
was first suggested in \cite{prl,Plerou,dublin,Cracow}. In this context $r_i^t$ are
daily (or higher frequency) returns of $N$ different stocks over a time period
of size $T$. As discussed above, even in the absence of any correlation between stocks, we expect 
the eigenvalues of an empirical correlation matrix, determined over
a finite time interval, to be non trivial. As a consequence, distinguishing between 
noise and genuine information in the empirical density of states is a subtle matter. 
For example, the diagonalisation of the correlation matrix of, say, 450
stocks computed over a bit more than 4 years (1125 trading days) reveals one very large
eigenvalue corresponding to a roughly uniform eigenvector, corresponding to
the ``market mode'', and a handful of other large eigenvalues that can be
seen to correspond to large sectors of economic activity. Smaller
eigenvalues,
however, form a blob around $\lambda=1$. Is this blob well described 
by the Mar\v{c}enko-Pastur distribution? In order to answer this question 
one has to rescale the small (supposedly ``pure noise'') 
empirical eigenvalues in order to have ${\overline \lambda}=1$ and compare 
to a Mar\v{c}enko-Pastur distribution (for which 
one has by definition ${\overline \lambda}=1$). Different results 
are obtained depending on the number of largest eigenvalues that are 
considered meaningful. They are shown in Figs. 3 and 4 below. 
The agreement between empirical data
and the Mar\v{c}enko-Pastur distribution is seen to be acceptable, although some systematic 
deviations are observed \cite{prl,Cracow}.
Should these deviations then be interpreted as the presence of true
economic information hidden in the noise band, or are we missing an important effect?
The phenomenology of financial markets suggests that a more faithful model 
consists in assuming that all individual stock returns are impacted by the same, time
dependent scale factor $\sigma_t$ that represents the ``market
volatility'': $r^t_i = \sigma_t \eta^t_i$, 
where $\eta_t^i$ are zero mean, unit variance Gaussian variables, and
$\sigma_t$ is a random variable.
From empirical studies, we know that the $\sigma_t$'s have long range
temporal correlations but for the purpose of the present study we only have
to specify the marginal distribution $P(\sigma)$. One possible choice that
matches quite well the data is to choose Eq. (\ref{defstudent})
with $\mu\simeq 3-5$. This corresponds to a multivariate Student distribution for the
returns, discussed for example in \cite{book,Sornette}.
This is the model we will consider more specifically below. Another possible choice,
inspired by the multifractal random walk model \cite{MRW}, is a log-normal
distribution for $P(\sigma)$.

\section{Wishart-Student matrices}

\subsection{Density of states of Pearson and Maximum Likelihood Estimators}

The first question we address is the generalisation of the Mar\v{c}enko-Pastur
spectrum for multivariate Student variables, when the Pearson estimate
Eq.\,(\ref{defE}) is used to determine the empirical correlation matrix.
The computation of the density of states (DOS) can be straightforwardly
performed using free random matrices techniques \cite{Verdu}. The trick is to use the so
called Blue function which is the inverse of the resolvent $G$: $B(G(z))=z$. The quantity
$B(x)-1/x$ is called the R-transform of $G$ and under certain hypotheses,
obeyed by any elliptic Wishart ensemble such that $C_{ij}=\langle \eta_i
\eta_j \rangle= \delta_{ij}$, is known to be additive \cite{Verdu}.\footnote{Note that
the ensemble we consider is {\it different} from the ensemble studied in
\cite{Burda,Bohigas} where the random volatility is time independent. In
this case, the average DOS is simply a convolution of the scaled
Mar\v{c}enko-Pastur
result with $P(\sigma)$. In this case, the result is clearly not
self-averaging.}

Since any elementary matrix $\sigma_t^2 \eta_i^t \eta_j^t$ is a projector,
its resolvent is simply:
\be
G_t(z)=\frac 1 N \left(\frac{1}{z-\frac{\overline \mu}{Qs_t}}+\frac{N-1}{z}
 \right)
\ee
where $Q=T/N$ and $\sigma_t^2 \equiv \overline \mu/s_t$ (henceforth, $\overline{\mu}=\mu/2-1$), 
such that $P(s)=s^{\mu/2-1}e^{-s}/\Gamma(\mu/2)$ in the
Student case. We have used that in the large $N$ limit $\sum_i
(\eta_i^t)^2/N=1$.
Inverting the resolvent at leading order we find:
\be
B_t(x)=\frac 1 x+\frac{\frac{\overline\mu}{Qs_t}}{N(1-\frac{x\overline\mu}{Qs_t})}
\ee
Using the additive properties of the R-transform we finally find the Blue
function for $\mathbf E$:
\be
B(x)=\frac 1 x+\frac 1 T
\sum_t\frac{\frac{\overline\mu}{s_t}}{(1-\frac{x\overline\mu}{Qs_t})}=
\frac 1 x+\int ds P(s) \frac{\frac{\overline\mu}{s}}{(1-\frac{x\overline\mu}{Qs})}
\ee
where the second identity is due to the large $T,N$ limit at fixed $Q$.

The relation between the resolvent and the density of states is
$G(\lambda-i\epsilon)=G_R(\lambda)+i\pi \rho(\lambda)$, where $G_R$ is the
real part of the resolvent.
Inverting this relation we find two coupled equations on $G_R, \rho$:
\begin{eqnarray}\label{doseq1}
\lambda&=&\frac{G_R}{G_R^2+\pi^2 \rho^2}+\int ds P(s) \frac{\overline\mu (s-\overline\mu
G_R/Q)}{(s-\overline\mu G_R/Q)^2+\overline \mu ^2\pi^2\rho^2/Q^2}\\
0&=&\rho\left(-\frac{1}{G_R^2+\pi^2\rho^2}+\int ds P(s)
\frac{\overline\mu^2/Q}{(s-\overline\mu G_R/Q)^2+\overline \mu ^2\pi^2\rho^2/Q^2}
\right)\label{doseq2}
\end{eqnarray}
Note that these equations are actually valid for any $P(\sigma)$ (when tails are
not too heavy).
The last equation of course always admits the $\rho=0$ solution. At very
small $\lambda$, the solution
of these equations is indeed $\rho=0$. The corresponding $G_R$ solves the
equation
\begin{equation}\label{realeq}
\lambda=\frac{1}{G_R}+\int ds P(s) \frac{\overline\mu}{s-\overline\mu G_R/Q}
\end{equation}
The RHS is well defined only for negative $G_R$, it goes to zero for very
large and negative $G_R$,
it goes to minus infinity at $G_R=0^-$ and it has a maximum somewhere in
between. The maximum of the RHS corresponds
to the largest value of $\lambda$ for which there is a real solution, i.e.
to the left edge of the DOS.
It can be determined obtaining the value of $G_R$ where the RHS of the
previous equation has a maximum:
\be
1=\int ds P(s) \frac{\overline\mu^2 G_R^2/Q}{(s-\overline\mu G_R/Q)^2},
\ee
and plugging this value into (\ref{realeq}). When $P(s)$ extends to $s=0$,
as is the case for Student variables, there is 
no real solution for larger values of $\lambda$. This implies that the DOS
has no right edge in that case.
In order to determine the DOS right tails we focus on the large $\lambda$
limit of eqs (\ref{doseq1},\ref{doseq2}).
It is easy to check that Eq. (\ref{doseq1}) is solved in the large $\lambda$
limit by $G_R \approx 1/\lambda$ if
$\rho/G_R$ goes to zero. We will check that this is indeed the case after
having determined $\rho(\lambda)$. In order to do so,
we analyze Eq. (\ref{doseq2}) assuming  $\rho/G_R \rightarrow 0$. In this
case
the integral can be computed exactly and one gets (for large $\lambda$):
\be
\rho(\lambda)\simeq \frac{{\overline \mu}^{\mu/2}}{\Gamma(\mu/2) Q^{\mu/2-1}}\frac{1}{\lambda^{1+\mu/2}}.
\ee
This result abides our
initial assumption $\rho/G_R = \rho \lambda \rightarrow 0$ when $\lambda \to \infty$.  It can be
interpreted in terms of rare events. Indeed the distribution of $\sigma$ has
exactly the same power
law tail. A very large $\sigma^*$ on a given day, $t$, leads to an
quasi-eigenvalue $\lambda \simeq \sigma^{*2}/Q$ plus subleading
contribution. As a consequence, writing (for $s<<1,\lambda>>1$) $T P(s) ds= N \rho(\lambda) d\lambda$ allows one 
recover precisely the
left tail of the DOS. As expected, provided $\mu > 2$, the power-tail disappears in the limit $Q \to \infty$.

\begin{figure}
\includegraphics[angle =-90, scale=0.35]{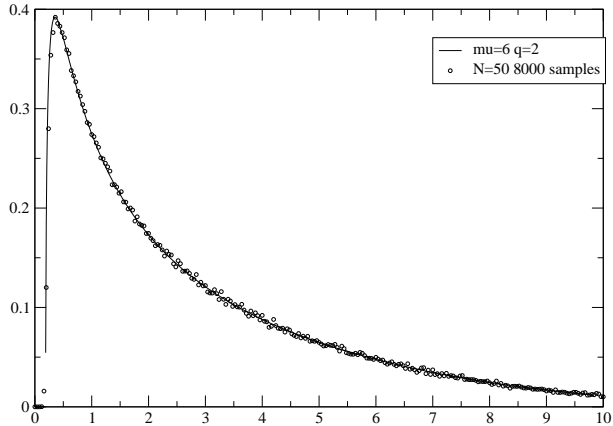}
\caption{\label{fig-dos} Comparison between the Wishart Student
distribution for $Q=2$ and $\mu=6$, solution of the previous equations,
and the numerical result obtained for $N=50$ and 8000 samples.}
\label{fs}
\vspace{-0.5cm}
\end{figure}

The Wishart-Student distribution for $Q=2$ and $\mu=6$, solution of the
previous equations, is plotted in Fig. \ref{fig-dos} and compared
to a numerical result obtained for $N=50$ and 8000 samples. The agreement
is excellent.

\begin{figure}
\includegraphics[angle =-90, scale=0.35]{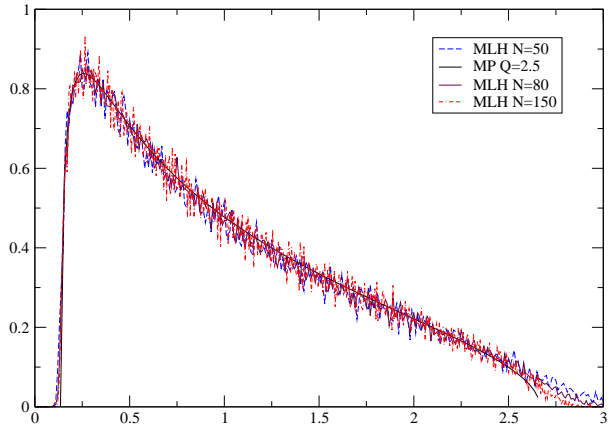}
\caption{\label{testMLH} Comparison between the numerical solution of 
the self-consistent equation (\ref{MLH-eq}) for $N=50,80,150$ and $Q=2.5$ (500 samples) and the 
Mar\v{c}enko-Pastur distribution. The agreement is excellent. Notice that the blurring of the MP right edge 
is a finite size effect, as shown by the evolution from $N=50$ to $N=150$.}
\label{fs}
\vspace{-0.5cm}
\end{figure}

We should however point out that the Pearson
estimator is not, in the case of Student variables, the maximum likelihood
estimator of $\bf C$. The formula for this estimator was worked out in
\cite{book} and is given instead by the solution of:
\be\label{MLH-eq}
E^*_{ij} = \frac{N+\mu}{T} \sum_t \frac{r_i^t r_j^t}{\mu + \sum_{mn} r_m^t
(E^{*-1})_{mn} r_n^t}
\ee
This reproduces the usual Pearson estimate when $\mu \to \infty$ at fixed $N$.
We are interested in the other limit $N,T \to \infty$  at fixed $\mu$. In this case $\mu$ can be dropped everywhere. 
As a consequence, the previous equation simplifies into:
\be
E^*_{ij} = \frac{1}{T} \sum_t \frac{\eta_i^t \eta_j^t}{N^{-1}\sum_{mn} \eta_m^t
(E^{*-1})_{mn} \eta_n^t}
\ee
Furthermore, in the large $N,T$ limit at fixed $Q$ the denominator in the RHS is expected to become
self-averaging and independent of $t$ at leading order. Since the above equation only fixes $E^*$ up to
an arbitrary multiplicative constant, we fix the value of the denominator to unity, which we can then verify self consistently. 
Since the $\eta_i^t$ are Gaussian random variables with unit variance, the maximum likelihood estimator of $\bf C$ 
is a Wishart matrix, and the eigenvalue spectrum is again given by the Mar\v{c}enko-Pastur distribution.
In order to check that the denominator is indeed equal to unity, we break $E^*$ into two contributions: 
$E^*_1+1/(T)\eta_i^t \eta_j^t$, where $E^*_1$ is the part of the Wishart Matrix independent of the $\eta_i^t$. 
Expanding the expression for the denominator in powers of $1/(T)\eta_i^t \eta_j^t$, one finds
at leading order in N:
\be
(\Tr (E^*)^{-1})/N \frac{1}{1+\Tr (E^*)^{-1}/(Q)}
\ee
where we have used that in the large $N,T$ limit $\Tr E^*/N\approx \Tr E^*_1/N$ and $\sum_i (\eta_i^t)^2/N=1$.
Recalling that the trace of the inverse of a Gaussian Wishart correlation matrix equals $Q/(Q-1)$ one can 
straightforwardly verify that the denominator is indeed equal to one.  
We have tested our result by numerical simulation by solving numerically 
the self-consistent equation (\ref{MLH-eq}) for $N=50,80,150$ and $Q=2.5$. The density 
of states, averaged over $500$ samples, is compared to the Mar\v{c}enko-Pastur distribution in Fig. \ref{testMLH}. 
The agreement is excellent, thus confirming our analytical result. 

This result is quite remarkable, especially from the point of view of cleaning noisy correlation matrices. 
Using the Maximum Likelihood Estimate of $\bf C$ one improves a lot the estimator; it allows 
to remove completely the effect of the noise due to $\sigma$ fluctuations, in particular 
to cut the noisy power-law tails of the Wishart-Student distribution.

At this stage, it is interesting to discuss the generality of the above results. 
First, remark that the DOS could also be
computed for an arbitrary correlation true matrix $C_{ij}=\langle \eta_i \eta_j \rangle$
using free random matrix theory and the so-called S-transform: since in
this case ${\bf E} = \sqrt{\bf C} {\bf E}_{WS} \sqrt{\bf C}$, where
${\bf E}_{WS}$ is a Wishart-Student empirical matrix considered above, the
eigenvalues of ${\bf E}$ will be the same as those of the product
${\bf C} {\bf E}_{WS}$. The spectrum of $\bf E$ can be computed from the
S-tranform of both $\bf C$ and ${\bf E}_{WS}$, extending the classical
result for Wishart matrices. Furthermore, our derivation of Eqs. (\ref{doseq1},\ref{doseq2})
does not require  Gaussian $\eta_i$s but only that  
$\sum_i (\eta_i^t)^2/N=1$ for large $N$. Finally, although we focused on a particular shape 
$P(\sigma)$, Eqs. (\ref{doseq1},\ref{doseq2}) generalizes straightforwardly to any $P(\sigma)$
with not too heavy tails, for example a log-normal distribution. 

\subsection{Applications to financial data}

\begin{figure}
\includegraphics[angle =-90, scale=0.35]{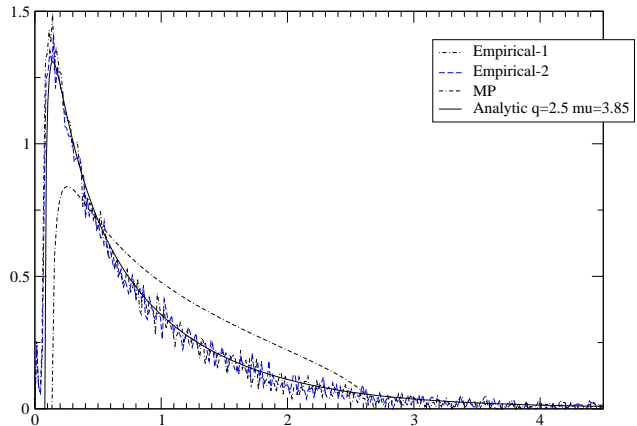}
\caption{\label{Emp1} Comparison  between the analytical results, Student (for $\mu=3.85$) and 
Gaussian Wishart, and the empirical data.}
\vspace{-0.5cm}
\end{figure}

\begin{figure}
\includegraphics[angle =-90, scale=0.35]{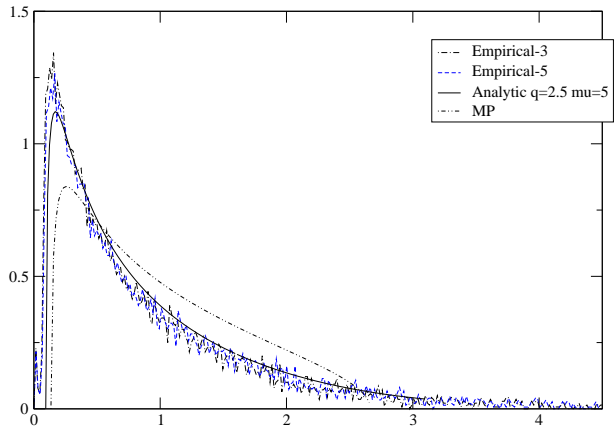}
\caption{\label{Emp2}  Comparison  between the analytical results, Student (for $\mu=5$) and 
Gaussian Wishart, and the empirical data.}
\vspace{-0.5cm}
\end{figure}

In the following we compare the Wishart-Student distribution to empirical data.
We have considered the daily returns of 450 stocks of SP-500
from 2003 to 2007 and computed the empirical (Pearson) correlation matrices ${\bf E}$ for $Q=2.5$.
The resulting average density of states is compared to the Mar\v{c}enko-Pastur
DOS and the Wishart-Student DOS for $\mu=4$ and $\mu=5$. 
Note that we renormalized the empirical DOS by $1-\sum_k^{K_m}\lambda_{k}/N$ 
where $k$ runs over the indexes of the largest $K_m$ eigenvalues.  
As discussed in the introduction, this is done in order to subtract a clearly non random 
contribution. Although the first few eigenvalues are certainly non-random the precise
choice of $K_m$ is a subtle matter. Different analysis suggest that $K_m\approx 10$ \cite{prl}. 
In the following we have chosen to determine $K_m$ directly from the Wishart-Student
DOS: we determine the values of $\lambda_{0.5},\lambda_{0.9}$ such that the probability that all 
sampled eigenvalues are less than these values is either 0.5 or 0.9 (of course assuming that 
the underlying distribution is Student). All eigenvalues that are larger than these cut-offs 
are assumed to be meaningful.  $\lambda_{0.5},\lambda_{0.9}$ depend on $\mu$; furthermore 
the determination of $K_m$ can be altered by other effects not taken into account in our simple model. 
Therefore we have verified that our results are stable when $K_m$ is between $2$ and $10$.
In Fig. \ref{Emp1} and \ref{Emp2} we show the comparison between the analytical results, Student and 
Gaussian Wishart, and the empirical data. We have used 20 samples corresponding to a sliding average 
with step of $15$ days. 
The two figures correspond respectively to the $\mu=3.85$ and $\mu=5$ case. In the first case we renormalized
the empirical data considering the first or the first two eigenvalue as meaningful.
In the second case we considered the first three or the first five as meaningful.
 
The agreement with the Wishart-Student DOS is surprisingly good. The optimal value of $\mu$ appears to be close to
$\mu=3.85$ in agreement with the value of $\mu$ obtained from the marginal distribution of
daily returns ($\mu \approx 3.85$, see \cite{book}).  As stressed above these results are not affected much as as long 
as $K_m$ is between $2$ and $10$. 

In order to test directly the hypothesis that the returns are multivariate Student variables, we have 
repeated the same analysis with daily returns scaled by a proxy of the instantaneous volatility, namely:
\be
\eta_i^t = \frac{r_i^t/\sigma_i^t}{\sqrt{N^{-1} \sum_{j=1}^N (r_j^{t}/\sigma_j^t)^2}}
\ee
and studied the eigenvalue spectrum of ${\widehat E}_{ij} = 1/T \sum_t \eta_i^t \eta_j^t$, again subtracting the top 
eigenvalues. Note that we have normalized each return by $\sigma_i^t=\sum_{t'\neq t} (r_i^t)^2/T$ as discussed in \cite{book}.

\begin{figure}
\includegraphics[angle =-90, scale=0.35]{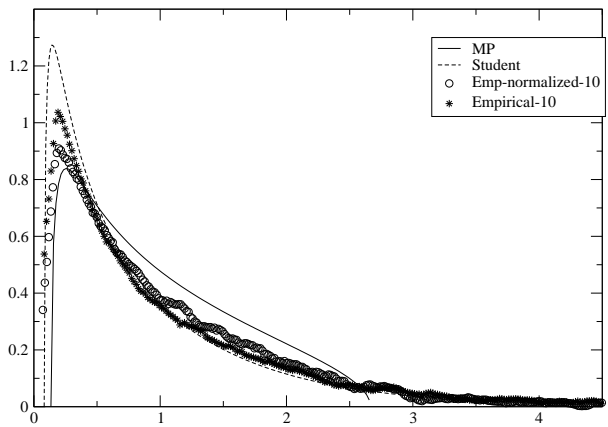}
\caption{\label{norm2} Comparison between the standard DOS of Wishart correlation matrices obtained from empirical data (Empirical-10) and the one obtained renormalizing the returns by a proxy of the daily volatility (Emp-normalized-10).
Both are rescaled using $K_m=10$ ($Q=2.5$). MP and Student are the analytical Mar\v{c}enko-Pastur and 
Student ($\mu=4$) results.}
\vspace{-0.5cm}
\end{figure}

In theory, the resulting spectrum should now be well fitted by the Mar\v{c}enko-Pastur distribution.
As shown in Fig. \ref{norm2}, the spectrum of $\bf{\widehat E}$ indeed moves closer to the Mar\v{c}enko-Pastur result but is still
distinctly different. However, the rescaled returns $\eta_i^t$ are still found have fat-tails; a possible
explanation is that the volatility of a given stock fluctuates not only through a common (market) factor $\sigma_t$ but
also through common sectorial volatilities, which could explain the deviation from Mar\v{c}enko-Pastur. Another possibility is of course 
that there are still non-trivial eigenvectors within the ``noise blob''. We leave the detailed study of this question for further studies.

\section{Kullback-Leibler entropy}

The Kullback-Leibler (KL) entropy allows one to measure the distance
between two
probability distributions and is defined in the following way \cite{KL}:
\be
S(2|1)=\int dx P_1(x) \log \left(\frac{P_1(x)}{P_2(x)}\right)
\ee
It is easy to show that this entropy is semi-positive definite using that
$S_{21}=-\langle \log P_2/P_1 \rangle_{P_1} \ge -\log \langle P_2/P_1
\rangle_{P_1}=
-\log \langle 1 \rangle_{P_2}=0$. Furthermore its minimum is reached for
$P_1=P_2$. As a consequence it
is a possible measure how much the distribution $P_2$ differs from $P_1$.
Note however that it is an
asymmetric measure: $S(1|2) \neq S(2|1)$.

Tumminello et al. \cite {TLM} computed the KL entropy for two multivariate
Gaussian
distributions with correlation matrices ${\bf C}_{1,2}$ and found:
\be
S({\bf C}_2;{\bf C}_1)=-\frac 1 2 \left[\Tr\log {\bf C}_2^{-1}{\bf
C}_1+\Tr({\bf C}_2^{-1}{\bf C}_1)
-\Tr {\mathbf 1}\right]
\ee
From this expression, they established that for $N$-multivariate Gaussian
variables (stock returns) with true correlation
matrix ${\bf C}$, the following holds:
\begin{itemize}
\item The average of $S({\bf E};{\bf C})$ is independent of ${\bf C}$ for
all $N$ where ${\bf E}$ is the Pearson estimator of
${\bf C}$ on a $T$ long
time series: $\langle S({\bf E};{\bf C}) \rangle = Z(N,T)$
\item The average of $S({\bf E}_1;{\bf E}_2)$ is also independent of the
true correlation ${\bf C}$,
where ${\bf E}_1,{\bf E}_2$ are the empirical correlations corresponding to
two independent
realisations of the multivariate process described by $\bf C$: $\langle
S({\bf E_1};{\bf E_2}) \rangle = Z'(N,T)$
\end{itemize}
This is a very interesting remark because in cases where the {\it a priori}
distribution is
indeed Gaussian one can judge the relative performance of different
cleaning procedures of $\bf E$ without
knowing the true correlation matrix $\bf C$, by computing $S({\bf E}; {\bf
E}_{cl})$ where ${\bf E}_{cl}$ is the
cleaned empirical correlation matrix. The above two values $Z,Z'$ provide
interesting benchmark values for $S$. The best one can do is to recover the true correlation
matrix: ${\bf E}_{cl}={\bf C}$, so $S({\bf E}; {\bf E}_{cl}) < Z$ means that some noise remains in the
cleaned matrix, while $S({\bf E}; {\bf E}_{cl}) > Z$ means that the cleaning is too violent and 
introduces some distortion. On the other hand, the most
trivial cleaning procedure is doing nothing: ${\bf E}_{cl}={\bf E}$.
Therefore, considering two independent realizations, a good cleaning procedure should be such that 
$S({\bf E}_1; {\bf E}_{2,cl}) \leq Z'$. 
Furthermore, $Z'$ is also interesting to estimate the reproducibility of the filtering procedure 
by comparing it to $S({\bf E}_{1,cl}; {\bf E}_{2,cl})$, see the discussion in \cite{TLM}. 
We also remark also that
the value of the KL entropies cited above (divided by $N$) is self-averaging in the limit
of a very large number of stocks $N$.

As discussed above, the distribution of stock returns is not Gaussian but
rather multivariate Student with an
exponent $\mu \approx 4-5$. [Note that the marginal distribution of single
stock returns is then also found to be a Student-t
distribution]. In order to apply the above ideas to real data, we need to
extend the results of Tumminello et al. \cite{TLM}
to multivariate elliptic distributions, parameterized by an arbitrary
distribution $P(s)$ of the inverse variance.
As explained above, the Gaussian case corresponds to $P(s)=\delta(s-\overline \mu)$ and
the Student case corresponds to $P(s)=s^{\mu/2-1}e^{-s}/\Gamma(\mu/2)$. 

\subsection{Kullback-Leibler entropy for elliptic laws}

In order to compute $S(2|1)$ for generic elliptic laws one has to compute:
\be
S_{12}=\int d{\bf x} P_1({\bf x}) \log \left(P_2({\bf x})\right),
\ee
where ${\bf x}$ is an $N$-dimensional vector.
The Kullback Leibler entropy will then be obtained as $S_{11}-S_{12}$.
Therefore all constant terms, i.e. independent of the correlation matrix
${\bf C}$, cancel between the two contributions.

A general expression for $S_{12}$ can be worked out using replicas:
\be
S_{12}=\lim_{n\rightarrow 0}\partial_n \int dx_1...dx_N \, P_1 (P_2)^n.
\ee
For any positive integer $n$ one can plug into the above equation the
general expression of multivariate elliptic laws and find the following
expression for $S_{12}$:
\be
\lim_{n\rightarrow 0}\partial_n \int dx_1...dx_N \, ds ds_1 ...ds_n
P(s) s^{N/2} \prod_{a=1}^n P(s_a) s_a^{N/2} \frac{\exp \left(-\frac 1 2
\sum_{i,j} x_i x_j
[s {\bf C}_1^{-1}+\sum_a s_a {\bf C}_2^{-1}]_{ij} \right)}{\sqrt{2
\pi}^{N(n+1)}
\sqrt{\det {\bf C}_1 \det ({\bf C}_2)^n}}
\ee
One can now integrate out the $x$ variables and get:
\be
S_{12}=\lim_{n\rightarrow 0}\partial_n \int ds ds_1 ...ds_n
P(s) \prod_{a=1}^n P(s_a) s_a^{M/2} \left(\frac{1}{\sqrt{2 \pi}^{N}
\sqrt{\det ({\bf C}_2)}} \right)^n\frac{1}{
\sqrt{\det [{\mathbf 1}+ {\bf C}_2^{-1} {\bf C}_1 \sum_a s_a/s]}}
\ee
Introducing the following identity in the previous expression:
\be
\int dy\delta\left(y-\sum_a s_a \right)=\int dy d\hat y
\exp(-iy\hat y +i\hat y \sum_a s_a),
\ee
one can finally make the analytic continuation to real $n$:
\be
S_{12}=\lim_{n\rightarrow 0}\partial_n \int ds dy d\hat y P(s)
\left(\frac{\int ds P(s) s^{N/2} e^{is\hat y}}{\sqrt{2 \pi}^{N}
\sqrt{\det ({\bf C}_2)}} \right)^n\frac{e^{-iy\hat y}}{
\sqrt{\det [{\mathbf 1}+ {\bf C}_2^{-1} {\bf C}_1 y/s]}}
\ee
It is now possible to differentiate with respect to $n$, take the limit $n \to 0$ and get the general
expression:
\be
S_{12}=-\frac 1 2 \Tr \log  {\bf C}_2+
\int ds dy d\hat y P(s)
\log \left(\frac{\int ds P(s) s^{N/2} e^{is\hat y}}{\sqrt{2 \pi}^{N}
\sqrt{\det ({\bf C}_2)}} \right)\frac{e^{-iy\hat y}}{
\sqrt{\det [{\mathbf 1}+ {\bf C}_2^{-1} {\bf C}_1 y/s]}}+K
\ee
where $K$ is a constant independent of the ${\bf C}$s.
The final expression for the Kullback-Leibler entropy is therefore:
\begin{eqnarray}
S(2|1)&=&-\frac 1 2 \Tr \log [{\bf C}_2^{-1}{\bf C}_1]-
\int ds dy d\hat y P(s)
\log \left(\int ds P(s) s^{N/2} e^{is\hat y} \right)\frac{e^{-iy\hat y}}{
\sqrt{\det [{\mathbf 1}+ {\bf C}_2^{-1} {\bf C}_1 y/s]}}+\nonumber\\
&&
\int ds dy d\hat y P(s)
\log \left(\int ds P(s) s^{N/2} e^{is\hat y}\right)\frac{e^{-iy\hat y}}{
(1+y/s)^{N/2}} \label{ge}
\end{eqnarray}
An important remark that will be very useful below is that this
expression can be written in general as $S(2|1)=\Tr f({\bf C}_2^{-1}{\bf
C}_1)$,
where $f$ is a function that depends on $P(s)$. In the following we will
apply
the general expression above to the Gaussian case, to check its validity,
and
to the Student case.
\subsubsection{Gaussian distribution}
Let us focus on the multivariate case when $P(s)=\delta(s-\overline \mu)$. In this case
the second term in the general expression above simplifies considerably. Up to 
a constant term that cancels out between the second and the third term 
and after rescaling $y, \hat y \rightarrow \hat y \overline \mu, y/\overline\mu$ we find:
\be
-i \int d\hat y dy \frac{\hat y e^{-i y \hat y}}{\sqrt{\det[{\mathbf 1}+
{\bf C}_2^{-1} {\bf C}_1 y]}}
\ee
Integrating over $\hat y$ one gets:
\be
\int dy \delta '(y)\frac{1}{\sqrt{\det[{\mathbf 1}+ {\bf C}_2^{-1} {\bf
C}_1 y]}}=\frac 1 2 \partial_y \left. \Tr \frac{{\bf C}_2^{-1} {\bf
C}_1}{{\mathbf 1}+ {\bf C}_2^{-1} {\bf C}_1 y}
\right|_{y=0}=\frac 1 2 \Tr [{\bf C}_2^{-1} {\bf C}_1]
\ee
Repeating the same procedure for the third term, we finally recover the
expression derived in \cite{TLM}:
\be
S(2|1)=\frac 1 2 \left[-\Tr\log {\bf C}_2^{-1}{\bf C}_1+\Tr({\bf
C}_2^{-1}{\bf C}_1)
-\Tr {\mathbf 1}\right],
\ee
which can indeed be obtained in a much simpler way. 
\subsubsection{Student distribution}
In order to proceed we have first to compute:
\be
\int ds P(s)s^N e^{i\hat y s}
\ee
which in the case of a Student distribution
$P(s)=s^{\mu/2-1}e^{-s}/\Gamma(\mu/2)$ reads:
\be
\frac {\Gamma(N+\mu/2)}{\Gamma(\mu/2)}
(1-i\hat y)^{-N-\mu/2}
\ee
We will use this identity to simplify the second term of (\ref{ge})
finding:
\be
\frac{N+\mu}{2}\int ds dy d\hat y P(s)
\log \left(1-i\hat y \right)\frac{e^{-iy\hat y}}{
\sqrt{\det [{\mathbf 1}+ {\bf C}_2^{-1} {\bf C}_1 y/s]}}+K'
\ee
In order to be able to integrate over $\hat y$ we apply the
identity $\log z =\int_0^{\infty}(e^{-t}-e^{-zt})/t$ to
$z=1-i\hat y$. Hence, we obtain:
\be
\frac{N+\mu}{2}\int ds dt P(s) \left(\frac{\exp(-t)}{t}-\frac{\exp(-t)}{t
\sqrt
{\det[{\mathbf 1}+ {\bf C}_2^{-1} {\bf C}_1 t/s]}} \right)
\ee
The double integral is dominated by small values of $t/s$ such that 
the leading contribution can be obtained expanding to the first order in 
$t/s$. This can be readily checked when ${\bf C}_2^{-1} {\bf C}_1=\mathbf 1$. 
Thus, in the large $N$ limit, one finds:
\be
\frac{N+\mu}{2}\int ds dt P(s) \left(\frac{\exp(-t)-\exp(-t-
t \Tr{\bf C}_2^{-1} {\bf C}_1/(2s))}{t} \right)
\ee
Using again the integral expression of the logarithm we find that the
second
term of (\ref{ge}) reads:
\be
\frac {N+\mu}{2}\int ds P(s)\log\left(1+\Tr{\bf C}_2^{-1} {\bf C}_1/2s\right)
\ee
Putting all pieces together we find finally the expression for the
Kullback-Leibler entropy for multivariate student distributions:
\be
S(2|1)=-\frac 1 2 \Tr \log [{\bf C}_2^{-1}{\bf C}_1]+
\frac {N+\mu}{2}\int ds P(s)\log\left(\frac{1+\Tr{\bf C}_2^{-1} {\bf C}_1/2s}
{1+N/2s}\right)
\ee
This expression allows one to recover the Gaussian case above in the limit $\mu/N \rightarrow \infty$, 
where $s \sim \mu/2 \gg N$. As a consequence one can expand the logarithm and find the previous expression 
for Gaussian distribution. The case of interest here is instead $\mu/N \ll 1$. In this case
the previous expression simplifies into:
\be
S(2|1)=-\frac 1 2 \Tr \log [{\bf C}_2^{-1}{\bf C}_1]+\frac N 2 \log\left(\Tr{\bf C}_2^{-1} {\bf C}_1/N\right)
\ee
One can change continuously from the Gaussian case to the Student case by tuning 
the parameter $x=N/\mu$ in the large $N,\mu$ limit. Gaussian and Student 
correspond respectively to $x=0,\infty$. 
Note that the final expression is independent of $\mu$ in the $x=\infty$ case, 
at least its leading contribution in $N$ which is what we are interested in.

\subsection{Applications}

Following \cite{TLM} we shall now compute $Z/N$ and $Z'/N$, i.e. $S_{KL}({\bf C}_1,{\bf C}_2)$
when ${\bf C}_1={\bf E}_1$ is the empirical correlation matrix generated
from the {\it a priori} correlation matrix ${\bf C}$ and ${\bf C}_2$ is
either equal to ${\bf C}$ or is another independent empirical correlation
matrix ${\bf E}_2$. The crucial results of \cite{TLM} is that these
expectation
values are independent of the true correlation matrix ${\bf C}$.
This is in fact true in the case of a general multivariate elliptic
distribution since
the final expression can be written as $\Tr f({\bf C}_2^{-1}{\bf C}_1)$. A
generic empirical correlation matrix ${\bf E}$
can indeed be written as: ${\bf E}=\sqrt{{\bf C}}{\bf E}_0\sqrt{{\bf C}}$
where ${\bf E}_0$ is a Wishart
correlation matrix of independent random variables. As a consequence the
contribution of ${\bf C}$ cancels out
from traces of powers of ${\bf E}_2^{-1}{\bf C}_1$ and one gets, for
example:
\be
S(2|1)=\Tr f({\bf E}_1^{-1}{\bf C})=\Tr f({\bf E}_0^{-1})
\ee
i.e.
as found in \cite{TLM}, the Kullback-Leibler entropy does not depend on the
{\it a priori} correlation matrix ${\bf C}$. However, its value depends on
$P(s)$.
In order to compare to real data we will compute explicitely these
entropies for Gaussian and Student distributions.
Both are expected to be self-averaging quantities in the large $N$ limit. 

\subsubsection{Gaussian}

Calling $\rho_{MP}(\lambda)$ the Mar\v{c}enko-Pastur density of states of the
empirical correlation matrices (with ${\bf C}={\bf 1}$) we find, from the
general expression above:
\be
\frac{S({\bf E};{\bf C})}{N}=\frac 1 2 \int d\lambda \rho(\lambda) [-\log
\lambda+1-\lambda]
\ee

In order to compute $S({\bf E}_1;{\bf E}_2)$ one has to calculate
$\Tr {\bf E}_1^{-1}{\bf E}_2/N$. This can be performed noticing that
the distribution law for these matrices is invariant under arbitrary
independent
rotations. Therefore we find
\be
\langle \Tr {\bf E}_1^{-1}{\bf E}_2/N\rangle=\frac 1 N \langle \sum_{a,b}
\frac{1}{\lambda_a} |\langle a | b \rangle |^2 \lambda_b \rangle=
\frac{1}{N^2}\langle \sum_{a,b}
\frac{1}{\lambda_a}\lambda_b \rangle=
\int d\lambda \rho_{MP}(\lambda) \lambda\int d\lambda \rho_{MP}(\lambda)\lambda ^{-1}
\ee
Hence, the final result is:
\be
\frac{S({\bf E}_1;{\bf E}_2)}{N}=-\frac 1 2 +\frac 1 2\int d\lambda
\rho_{MP}(\lambda) \lambda \, \int d\lambda \rho_{MP}(\lambda)
\frac{1}{\lambda}
\ee
It can be shown that these expressions coincide with the ones of \cite {TLM} 
in the large $N$ limit. 

\subsubsection{Student}

Calling now $\rho_S(\lambda)$ the density of states of the Wishart-Student 
matrices computed in section II above, we find:
\be
\frac{S({\bf E};{\bf C})}{N}=-\frac 1 2 \int d\lambda \rho_S(\lambda) \log
\lambda
+\frac 1 2 \log\left( \int d\lambda \rho_S(\lambda)\lambda \right)
\ee

Applying the same argument used in the Gaussian case, one also finds:
\be
\frac{S({\bf E}_1;{\bf E}_2)}{N}=\frac 1 2\log \left(\int d\lambda \rho_S
(\lambda)\lambda \right)+\frac 1 2 \log \left(\int d\lambda \rho_S
(\lambda)/\lambda \right)
\ee

The numerical values of these entropies, computed for different $Q=T/N$ and
$\mu$'s, are given in the Tables below.

\begin{center}
   \begin{tabular}{ l || c | c | c | c }
     \hline
     $\mu=3$  &  $Q=1.5$  &  $Q=2$ & $Q=3$ & $Q=5$ \\ \hline
     $Z/N$ & 0.645126 & 0.527893 & 0.409243 & 0.303255 \\ \hline
     $Z'/N$ &  0.990942 &  0.730459 & 0.519955 & 0.361961 \\
     \hline
   \end{tabular}
 \end{center}

\begin{center}
   \begin{tabular}{ l || c | c | c | c }
     \hline
      $\mu=4$ & $Q=1.5$  &  $Q=2$ & $Q=3$ & $Q=5$ \\ \hline
     $Z/N$ & 0.445103 & 0.336914 & 0.233323 & 0.149822 \\ \hline
     $Z'/N$ &  0.814573 &  0.568792 & 0.376484 & 0.23867 \\
     \hline
   \end{tabular}
 \end{center}

\begin{center}
   \begin{tabular}{ l || c | c | c | c }
     \hline
      $\mu=5$ &$Q=1.5$  & $Q=2$ & $Q=3$ & $Q=5$ \\ \hline
     $Z/N$ & 0.361844 & 0.263336 & 0.172947 & 0.10532 \\ \hline
     $Z'/N$ &  0.739387 &  0.502362 & 0.320584 & 0.193947 \\
     \hline
   \end{tabular}
 \end{center}

A comparison of these values to financial data, along the same line as \cite{TLM}, 
is left for a future work. Note that since the Maximum Likelihood Estimate of the 
a priori correlation matrix is a Wishart-Gaussian matrix in the large $N$ limit (see section II),
the values of $Z,Z'$ found by \cite{TLM} in the case of Wishart-Gaussian matrices turn out 
to be correct for the Maximum Likelihood Estimate of Student correlation matrices, as very recently 
found numerically by Tumminello et al. \cite{TLM2}.

\section{Conclusion}

In this work, we have studied in some details a new ensemble of random correlation matrices related
to multivariate Student (or more generally elliptic) random variables. We have found the exact density
of states for the Pearson estimate of the correlation matrix for uncorrelated variables, that generalizes
the Mar\v{c}enko-Pastur result. It would be interesting to know whether the joint distribution of eigenvalues 
can also be computed exactly in this case. We have shown that for the Maximum Likelihood estimator, the density of states
is still exactly given by the Mar\v{c}enko-Pastur distribution.

The comparison between the theoretical density of states in the Student case and empirical financial data is 
surprisingly good, in any case much better than the Mar\v{c}enko-Pastur result. However, we are still able to 
detect significant systematic deviations, which suggest the need of a richer, non-elliptic model for the joint distribution of
returns, or the presence of information carrying, low eigenvalues of the correlation matrix (or both).

Finally, we have computed explicitely the Kullback-Leibler entropies of empirical Student matrices, which are
found to be independent of the true correlation matrix, as in the Gaussian case. Using our result on the density of
states, we give the exact numerical value of the Kullback-Leibler entropies in various cases of interest.

\vskip 1cm

We thank Fabrizio Lillo for very useful discussions and for sending ref. \cite{TLM2} prior to publications, and the
organizers of the second Cracow meeting on Random Matrices for providing us with the opportunity to put this work together.

\end{document}